# CULTURAL ANTHROPOLOGY THROUGH THE LENS OF WIKIPEDIA – A COMPARISON OF HISTORICAL LEADERSHIP NETWORKS IN THE ENGLISH, CHINESE, JAPANESE, AND GERMAN WIKIPEDIA

Peter A. Gloor[1], Patrick de Boer[1], Wei Lo[2], Stefan Wagner[3], Keiichi Nemoto[4], Hauke Fuehres[5]
MIT[1], Zhejiang University[2], FHNW[3], Fuji Xerox[4], galaxyadvisors[5]
Cambridge USA[1], Hangzhou China[2], Brugg Switzerland[3], Tokyo Japan[4], Aarau Switzerland[5]
pgloor@mit.edu, pdeboer@mit.edu, spencer_w_lo@zju.edu.cn, stefan.wagner@fhnw.ch,
keiichi.nemoto@fujixerox.co.jp, hfuehres@galaxyadvisors.com

## ABSTRACT

In this paper we study the differences in historical worldview between Western and Eastern cultures, represented through the English, Chinese, Japanese, and German Wikipedia. In particular, we analyze the historical networks of the World's leaders since the beginning of written history, comparing them in the four different Wikipedias.

## INTRODUCTION

In this project we are using Wikipedias in different languages as a window into the "soul" of different cultures, replacing anthropological fieldwork with statistical analysis of the treatment given by native speakers of a culture to different subjects in Wikipedia.

One of the most popular categories in Wikipedia is the people pages, talking about the most important people of all ages. Wikipedians have put together "notability criteria" that clearly define if a person deserves inclusion into Wikipedia or not. In this paper we look at the most prominent people pages over all times in the English, Chinese, Japanese, and German Wikipedia, resulting in a comparison between the Western and Eastern worldview.

## PEOPLE NETWORK CONSTRUCTION

Our goal was to create a social network of all people that every lived, since the beginning of time. As a proxy, we only take people that made it into Wikipedia, fulfilling Wikipedia's notability criteria. As a second requirement, a link between two people can only exist if both of them were living at the same time. For each Wikipedia, we start with all pages tagged as "people pages", in the English Wikipedia this are for instance 800,000 pages. In the next step all people pages are dated, by extracting the dates of birth and of death of each individual. Moreover, the links originating and pointing to their Wikipedia page are gathered. Using this information, for each year through history, from 3000 BC to 1950 CE, a link network is calculated, as shown in figure 1. From all the links originating and pointing back to a particular people page, only the links to and from people living at the same time as the person discussed on that page are included.

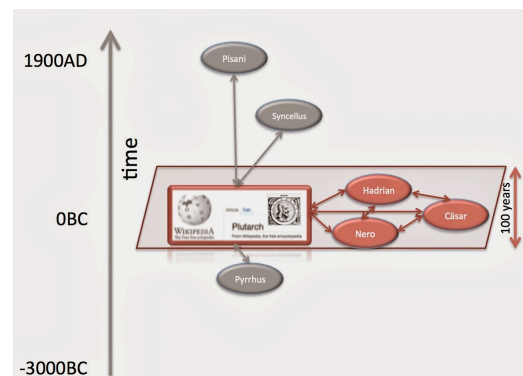

*Figure 1: Link construction among contemporary people pages*

For instance, in the graph shown in figure 1 above, from all the links to the page about Plutarch, only the links from and to Hadrian, Caesar, and Nero are kept, while the links to Pyrrhus, who died well before Plutarch was born, and the pages to medieval historian Syncellus and modern historian Pisani are ignored as well. Repeating this process leads to 4900 unique networks for the English Wikipedia (less for the Chinese and Japanese Wikipedia, as their history does not go as far back). For each of these networks, the most central people are determined using the PageRank algorithm. To get a second selection criteria among the influencers, their indegree, i.e. other people pages pointing back to them, is taken.

Figure 2 illustrates the Wikihistory application running in a Web browser that we developed based on the network extracted above. It shows the network of most influential people in the English Wikipedia in the year 0 – note that Jesus is not yet part of the network, he will show up in year 1.

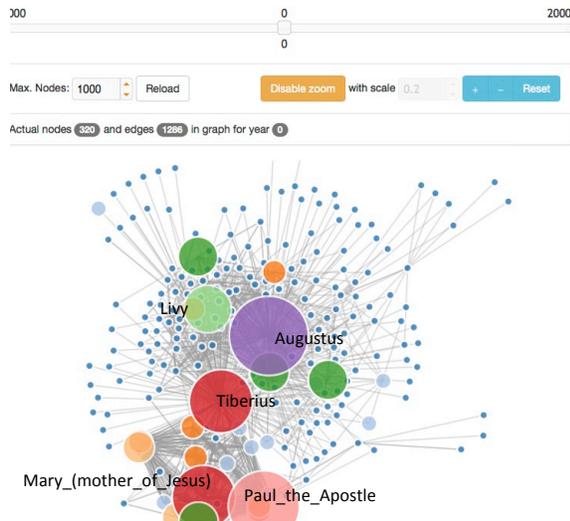

*Figure 2: Sample screen shot of Wikihistory at 0 BC*

## IT'S ALL ABOUT EMPERORS AND WARS

Who are the most important people of all times? The answer to this question is quite different in the US, the UK, and German-speaking countries than it is in China and Japan. Looking at the top ten and top 50 people lists also confirms that most English language Wikipedia editors come from the US and the UK, while Chinese language editors come from Taiwan, Hong Kong, and mainland China.

| English | Chinese | Japanese | German |
|---|---|---|---|
| George_W._Bush | Mao_Zedong | Ikuhiko_Hata | Adolf Hitler |
| William_Shakespeare | Yuan_Shikai | Tokugawa_Ieyasu | Johann Wolfgang von Goethe |
| Sidney_Lee | Jay_Chou | Toyotomi_Hideyoshi | Aristotle |
| Jesus | Oda_Nobunaga | Adolf_Hitler | Benedict XVI |
| Charles_II_of_England | Tokugawa_Ieyasu | Oda_Nobunaga | Plato |
| Aristotle | Emperor_Gaozong_of_Tang | Hirohito | Martin Luther |
| Napoleon | Cao_Cao | Tokugawa_Hidetada | Otto von Bismarck |
| Muhammad | Kangxi_Emperor | Tokugawa_Iemitsu | Johannes Paul II |
| Charlemagne | Emperor_Huizong_of_Song | Chiang_Kai-shek | Johann Heinrich Zedler |
| Plutarch | Yongle_Emperor | Tokugawa_Ienari | Johann Sebastian Bach |

*Table 1: Top ten most important people in 4 Wikipedias (most important at the top) Red denotes politicians, black religious leaders, blue scientists and artists*

As tables 1 and 2 illustrate, in China and Japan only famous warriors and politicians have a chance to make it into the top ten and top fifty – the East seems far less religious than the West – while the Western Wikipedias are more balanced with half of the top ten as well as the top fifty of all times being religious leaders or artists or scientists. Historians play a special role. Both Sidney Lee, a relatively minor Victorian professor of English and history, who wrote 800 biographies, and Ikuhiko Hata, a 19 century Japanese military biographer owe their prominent position to their prolific biography writing, as they get many backlinks from the references on the pages of contemporary politicians they wrote about.

|  | English | Chinese | Japanese | German |
|---|---|---|---|---|
| Politicians | 26 | 46 | 47 | 23 |
| Religious Leaders | 11 | 1 | 0 | 5 |
| Artists/ Scientists | 13 | 3 | 3 | 22 |
| Cultural Ingroup | 10 | 48 | 31 | 31 |

*Table 2: Distribution of different people categories in 4 Wikipedias among the top 50 people of all times*

The second striking difference comes from outgroup leaders included into the top 50. While the English Wikipedia includes 80% non-English leaders among the top 50, just two non-Chinese made it into the top 50 of the Chinese Wikipedia: Napoleon III and Tokugawa Ieyasu. The Japanese Wikipedia is slightly more balanced, with almost 40 percent non-Japanese leaders, half of them Chinese Emperors, the others people like Adolf Hitler, Plato, Cicero, and Augustus.

## CONCLUSIONS

The Internet enables researchers to more easily compile rankings of the most important world leaders of all times (Murray 2003, Hidalgo 2014). Our work is unique in that we extract language-specific rankings that allow us to compare the worldview for dozens of different cultures. Probing the historical perspective of many different language-specific Wikipedias gives an X-ray view deep into the historical foundations of cultural understanding of different countries.

## REFERENCES

Hidalgo, C. (2014) Pantheon: Mapping Historical Cultural Production. http://pantheon.media.mit.edu

Murray, C. (2003). Human accomplishment: The pursuit of excellence in the arts and sciences, 800 BC to 1950. HarperCollins.